\documentclass[prl,showpacs,twocolumn]{revtex4}
\usepackage{amsfonts}
\usepackage{amsmath}
\usepackage{amssymb}
\usepackage{graphicx}

\begin{document}

\title{Phase diffusion of a two-component Bose-Einstein condensates: exact and short-time solutions for arbitrary coherent
spin state}

\author{G. R. Jin, B. B. Wang, and Y. W. Lu}
\affiliation{Department of Physics, Beijing Jiaotong University,
Beijing 100044, China}

\begin{abstract}
We investigate phase diffusion of a two-component Bose-Einstein
condensates prepared initially in arbitrary coherent spin state
$|\theta_0,\phi_0\rangle$. Analytical expression of the
phase-diffusion time is presented for $\theta_0\neq\pi/2$ case. In
comparison with the symmetrical case (i.e., $\theta_0=\pi/2$), we
find that the diffusion process becomes slowly due to the reduced
atom number variance.
\end{abstract}
\pacs{03.75.Mn, 05.30.Jp,42.50.Lc} \maketitle

Phase diffusion of Bose-Einstein condensates (BECs)
\cite{Lewenstein,Wright,Imamoglu,Javanainen,Castin,Law98,Vardi},
aroused from atom-atom interactions destroys phase coherence, and
thus sets a limit to the applications of the condensates in
high-precision measurement and quantum information processing. The
phase diffusion is observable by measuring fringe visibility in
atomic interference experiments \cite{Orzel,Schumm,Chuu,Jo,Artur}.
For a two-component BECs prepared initially in a coherent spin state
(CSS) $|\theta_{0}, \phi_{0}\rangle$ with $\theta_{0}=\pi/2$ [see
Ref.~\cite{CSS}, or Eq.~(\ref{CSS})], it has been shown that the
single-particle coherence (or fringe visibility) decays
exponentially with the time scale $\chi t_{d}=(N/2)^{-1/2}$
\cite{Wright,Imamoglu,Javanainen,Castin}, dependent upon atom number
$N$ and the self-interaction strength $\chi \sim
(a_{11}+a_{22}-2a_{12})/2$. Here, $a_{ii}$ and $a_{ij}$ ($i\neq j$)
denote intra- and inter-species s-wave scattering lengths,
respectively. In this paper, we generalize previous works by
considering arbitrary initial CSS, i.e., $\theta_{0}\neq\pi/2$.
Exact and short-time solutions of the single-particle coherence are
obtained. In comparison with $\theta_{0}=\pi/2$ case, we find that
the diffusion becomes slowly due to relatively narrow atom number
variance.

Considering a two-component BECs with internal states $|1\rangle$
and $|2\rangle$ confined in a deep potential, we adopt single-mode
approximation (SMA)~\cite{TMA}, i.e., keeping the condensed-mode
wave function $\Phi_{0}(r)$ for the two components, so the total
system can be well described by second-quantized Hamiltonian ($\hbar
=1$):
\begin{eqnarray}
\hat{H}&=&\omega_{1}\hat{n}_{1}+\omega
_{2}\hat{n}_{2}+u_{12}\hat{n}_{1}\hat{n} _{2}\nonumber\\
&+&\frac{u_{11}}{2}(\hat{b}_{1}^{\dag
})^{2}(\hat{b}_{1})^{2}+\frac{u_{22}}{2}(\hat{b}_{2}^{\dag
})^{2}(\hat{b}_{2})^{2},  \label{H1}
\end{eqnarray}%
where $\hat{b}_{i}$, $\hat{b}_{i}^{\dag }$, and $\hat{n}_{i}$
($=\hat{b}_{i}^{\dag }\hat{b}_{i}$) are the annihilation, creation,
and number operators for the internal states $i=1,2$, respectively.
In addition, $\omega _{i}$ are single-particle energies, and
$u_{ij}=(4\pi a_{ij}/M)\int d^{3}r\vert \Phi _{0}(r)\vert^{4}$ are
atom-atom interaction strengthes. Particle number operator
$\hat{N}=\hat{n}_{1}+\hat{n}_{2}\ $ is a conserved quantity and is
set to the c number $N$. Hamiltonian (\ref{H1}) can be rewritten as
$\hat{H}=\delta \hat{J}_{z}+\chi \hat{J}_{z}^{2}$, where the
detuning $\delta =\omega _{2}-\omega _{1}+(u_{22}-u_{11})(N-1)/2$,
and the self-interaction strength $\chi =(u_{11}+u_{22}-2u_{12})/2$.
The spin operators $\hat{J}_{+}=(\hat{J}_{-})^{\dag
}=\hat{b}_{2}^{\dag }\hat{b}_{1}$ and
$\hat{J}_{z}=(\hat{n}_{2}-\hat{n}_{1})/2$, obeying SU(2) algebra.
The nonlinearity $\chi \hat{J}_{z}^{2}$ has been proposed to prepare
spin squeezed state \cite{Kitagawa,Jin07,Jin08} and also quantum
entangled state \cite{Wei,Liang}. Note that the model considered
here with nonzero $\delta$ can be also used to study the BECs in an
asymmetric double-well \cite{Wang}. Squeezing via coupling of the
BECs in a double-well potential with a cavity light field has been
investigated in Ref. \cite{Zhou}.

\begin{figure}[tbph]
\begin{center}
\includegraphics[width=3.1in]{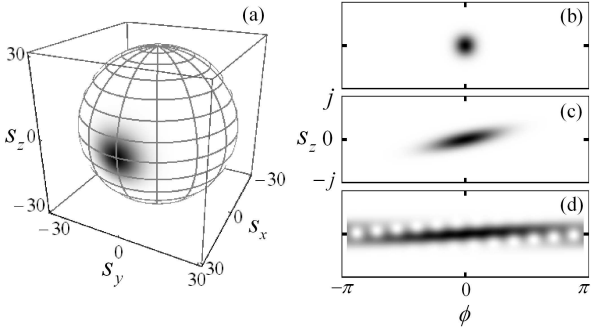}
\caption{ Husimi $Q$ function $Q(\theta, \phi; t)$ for initial CSS
$|\protect\theta_0=\pi/2, \phi_0=0\rangle$ on Bloch sphere (a), with
Bloch vector $\vec{s}=\langle\vec{J}\rangle=j(\sin\theta\cos\phi,
\sin\theta\sin\phi, \cos\theta)$ and $j=N/2=30$. The $Q$ functions
in the phase space $(\phi, s_{z})$ for various times: (b) $t=0$, (c)
$t=t_s=0.0764\chi^{-1}$, and (d) $t=2t_d=0.2651\protect\chi^{-1}$.
Time scale $\chi t_{s}\simeq 3^{1/6}(2j)^{-2/3}$ is for optimal spin
squeezing \cite{Kitagawa,Jin07,Jin08}, and $\chi t_{d}=j^{-1/2}$ is
phase-diffusion time
\protect\cite{Wright,Imamoglu,Javanainen,Castin}. The $Q$ function
is normalized by its maximum value.} \label{fig1}
\end{center}
\end{figure}

Besides the spin squeezing, mean-field interaction $\chi J_{z}^{2}$
also leads to phase diffusion in the two-component BECs
\cite{Lewenstein,Wright,Imamoglu,Javanainen,Castin,Law98,Vardi,Artur},
which can be illuminated schematically by Husimi $Q$ function
$Q(\theta ,\phi ;t)=|\langle \theta ,\phi |\Psi (t)\rangle |^{2}$,
where
\begin{equation}
|\theta, \phi\rangle=\exp\{i\theta (\hat{J}_{x}\sin \phi
-\hat{J}_{y}\cos \phi)\}|j,j\rangle \label{CSS}
\end{equation}%
is arbitrary CSS \cite{CSS} and $|\Psi (t)\rangle
=e^{-i\hat{H}t}|\Psi (0)\rangle $ is a state vector at any time $t$.
For an initial state $|\Psi (0)\rangle =|\theta _{0},\phi
_{0}\rangle $, analytic expression of the $Q$ function reads
\cite{CSS}
\begin{eqnarray}
Q(\theta, \phi; 0)&=&|\langle \theta, \phi|\theta _{0},\phi
_{0}\rangle|^{2}=\left[\frac{1+\cos\Theta}{2}\right]^{2j},
\end{eqnarray}%
where $\cos\Theta=\cos\theta\cos\theta_{0}+\sin\theta\sin\theta
_{0}\cos(\phi-\phi_{0})$. The $Q$ function can be plotted in
three-dimensional phase space (i.e., Bloch sphere)
\cite{Kitagawa,Jin07}, or alternatively, in a two-dimensional phase
space ($\phi, s_{z}$) via a mapping $s_{z}=j\cos\theta$
\cite{Artur,Jin08,Trimborn}. As shown in Fig.~\ref {fig1}(a) and
Fig.~\ref {fig1}(b), the density of $Q(\theta, \phi; 0)$ is
distributed isotropically, indicating the minimal uncertainty
relationship of the initial CSS. Under the government of nonlinear
interaction $\chi J_z^2$, the system will evolve into a spin
squeezed state \cite{Kitagawa} with anisotropic distribution of the
$Q$ function [see Fig.~\ref {fig1}(c)]. As time increases, the spin
system is over-squeezed \cite{Kitagawa,Shalm} and shows a spread of
the $Q$ function along the $\phi$ axis, which simulates an increased
relative phase fluctuation (i.e., phase diffusion) \cite{Artur}, as
shown in Fig.~\ref {fig1}(d).

%%%%%%%%%%%%%%%%%%%%%%%%%%%%%%%%%%%%%%fig.1
%%%%%%%%%%%%%%%%%%%%%%%%%%%%%%%%%%%%%%%%%%%%%%%%%%

\begin{figure}[htbp]
\begin{center}
\scalebox{0.6}{\includegraphics{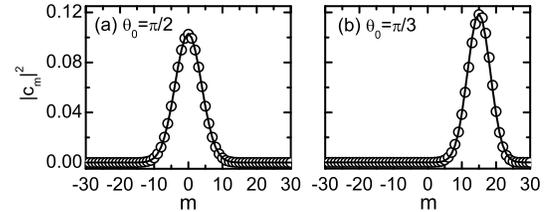}} \caption{Probability
distribution, $|c_m|^2$ as a function of $m$ for the initial CSS
$|\theta_0, \phi_0\rangle$. (a) $\theta_0=\pi/2$ and (b)
$\theta_0=\pi/3$. The empty circle is given by Eq.~(\ref{cm}), and
solid line is given by Eq.~(\ref{cma}). The parameter $j=N/2=30$.}
\label{fig2}
\end{center}
\end{figure}

It was shown that the most sensitive states to the diffusion are the
symmetrical CSS with $\theta _{0}=\pi /2$, corresponding to equal
populations between the two internal states \cite{Vardi}. The
phase-diffusion time scale is $\chi t_{d}=1/\sqrt{j}$ (with $j=N/2$)
\cite{Wright,Imamoglu,Javanainen,Castin}. Here, we generalize it for
arbitrary initial CSS
\begin{equation}
|\Psi(0)\rangle=\left\vert\theta_{0}, \phi_{0}\right\rangle
=\sum_{m=-j}^{j}c_{m}|j,m\rangle ,  \label{psit0}
\end{equation}%
where the amplitudes
\begin{eqnarray}
c_{m}&=&{2j \choose j-m}^{1/2}\cos^{j+m}(\frac{\theta _{0}}{2})
\sin^{j-m}(\frac{\theta _{0}}{2}) e^{i(j-m)\phi_{0}}. \label{cm}
\end{eqnarray}%
In single-particle picture, the CSS corresponds to all the atoms
occupying a superposed state \cite{Kitagawa}: $\cos (\theta
_{0}/2)|2\rangle +e^{i\phi _{0}}\sin (\theta _{0}/2)|1\rangle $,
where the polar angle $\theta _{0}$ and the azimuth angle $\phi
_{0}$ determine atom population and the relative phase between the
two internal states, respectively. In Fig.~\ref{fig2}, we plot
$|c_{m}|^{2}$ as a function of $m$. One can find that the
probability distribution of the CSS can be treated as a Gaussian
wave packet [see below Eq. (\ref{cma})]. At any time $t$, the spin
system evolves into
\begin{equation}
|\Psi (t)\rangle =\sum_{m=-j}^{j}c_{m}e^{-i(\delta m+\chi
m^{2})t}\left\vert j,m\right\rangle ,  \label{psit}
\end{equation}%
i.e., a superposition of atomic number state $|j,m\rangle
=|j-m\rangle _{1}|j+m\rangle _{2}$. Due to the presence of atom-atom
interaction ($\chi \neq 0$), each number state has different phase
evolution rate, which in turn lead to collapse and revival of the
Rabi oscillation, a phenomenon that is well-known in quantum optics
\cite{TMA}. Similar effect has been investigated in two-component
BECs \cite{Kuang}, atom-molecular BECs
\cite{Javanainen99,Jin05,Cai}, and exciton emission
\cite{Jin04,Glazov,Weibin}.

Phase diffusion of the BECs considered here is in fact collapse of
the first-order temporal correlation function \cite{Imamoglu,Vardi}:
\begin{equation}
g_{12}^{(1)}=\frac{\vert \rho _{12}^{(1)}\vert}{\sqrt{\rho
_{11}^{(1)}\rho _{22}^{(1)}}}\equiv \frac{\vert \langle
\hat{J}_{+}\rangle\vert }{\sqrt{j^{2}-\langle \hat{J}_{z}\rangle
^{2}}}, \label{g12}
\end{equation}%
where $\rho _{kl}^{(1)}=\langle \hat{b}_{k}^{\dag
}\hat{b}_{l}\rangle /N$ with $k,l=1,2$ are the elements of the
single-particle density matrix \cite{Vardi}. The coherence
$g_{12}^{(1)}$ is observable in experiments by extracting the
visibility of the interference fringes
\cite{Orzel,Schumm,Chuu,Jo,Artur}. The expectation value $\langle
\hat{J}_{z}\rangle=\langle \Psi(t)|\hat{J}_{z}|\Psi(t)\rangle$,
reads
\begin{equation}
\langle\hat{J}_{z}\rangle=\sum_{m=-j}^{j}m|c_{m}|^{2}=j\cos\theta
_{0}, \label{Jz}
\end{equation}%
which is a constant for a fixed polar angle of the initial CSS
$\theta_0$. This is because of conserved operator $\hat{J}_z$ with
respect to the Hamiltonian $\hat{H}$. Similarly, we obtain the
expectation value $\langle \hat{J}_{z}^{2}\rangle =j^{2}\cos
^{2}\theta _{0}+(j/2)\sin ^{2}\theta _{0}$, and%
\begin{eqnarray}
\langle \hat{J}_{+}\rangle&=&e^{i(\phi _{0}+\delta t+\chi t)}\cot
\frac{\theta _{0}}{2}\sum_{m=-j}^{j}(j-m)\left\vert c_{m}\right\vert
^{2}e^{2im\chi t}  \nonumber \\
&=&je^{i(\phi _{0}+\delta t)}\sin \theta _{0}\left[ \cos \chi
t+i\cos \theta _{0}\sin \chi t\right] ^{2j-1},  \label{J+}
\end{eqnarray}%
where we have used the relation:
$c_{m+1}=(j-m)^{1/2}(j+m+1)^{-1/2}\cot (\theta _{0}/2)e^{-i\phi
_{0}}c_{m}$. Inserting Eq. (\ref{Jz}) and Eq. (\ref{J+}) into Eq.
(\ref{g12}), we further obtain the exact solution of the coherence
\begin{equation}
g_{12}^{(1)}(t)=\left[1-\sin ^{2}\left( \theta _{0}\right) \sin
^{2}\left( \chi t\right) \right] ^{j-1/2},  \label{g12_exact}
\end{equation}%
which shows a decay of the coherence, i.e., phase diffusion. Such a
dephasing process depends sensitively on the self-interaction
strength $\chi $ and the polar angle $\theta _{0}$ of the initial
CSS; while the detuning $\delta$ and the azimuth angle $\phi _{0}$
gives \textit{vanishing} contribution to the coherence.

%%%%%%%%%%%%%%%%%%%%%%%%%%%%%%%%%%%%%%%%%%%%%%%%%%
%%%%%%%%%%%%%%%%%%%%%%%%%%%%%%%%%%%%%%%%%%%%%%%%%%

Time scale of the diffusion is determined by short-time behavior of
the coherence. Following Imamo\={g}lu et al. \cite{Imamoglu}, we
treat the initial CSS as a Guassian wave-packet with its peak
located at $m\simeq \langle \hat{J}_{z}\rangle =j\cos \theta_0$,
namely
\begin{equation}
\vert c_{m}\vert^{2}\simeq \frac{1}{[ 2\pi ( \Delta \hat{J}_{z})
^{2}]^{1/2}}\exp\left[-\frac{( m-\langle \hat{J}_{z}\rangle)
^{2}}{2( \Delta \hat{J}_{z})^{2}}\right], \label{cma}
\end{equation}%
where the width
\begin{equation}
(\Delta \hat{J}_{z})^{2}=\langle \hat{J}_{z}^{2}\rangle -\langle
\hat{J}_{z}\rangle ^{2}=(j/2)\sin ^{2}\theta _{0}. \label{variance}
\end{equation}
Due to conserved particle number $N$ ($=2j$),
$\hat{n}_{1}=j-\hat{J}_z$ and $\hat{n}_{2}=j+\hat{J}_z$, and thus
atom number variances $(\Delta \hat{n}_{1})^{2}=(\Delta
\hat{n}_{2})^{2}=(\Delta \hat{J}_{z})^{2}$ \cite{Jin08}.

\begin{figure}[htbp]
\begin{center}
\scalebox{0.5}{\includegraphics{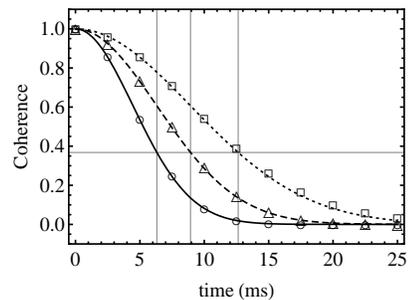}} \caption{Time evolution
of the first-order temporal coherence $g_{12}^{(1)}(t)$ for various
polar angles: $\theta_0=\pi/6$ (dotted line, squares), $\pi/4$
(dashed line, triangles), and $\pi/2$ (solid line, circles). The
empty squares, triangles, and circles are given by
Eq.~(\ref{g12_exact}) for corresponding $\theta_0$'s, while the
curves are given by Eq.~(\ref{Imamoglu}). Horizontal grid line gives
the value $1/e$, vertical lines denote the diffusion times $t_d$ for
different $\theta$'s. Other parameters are $N=60$,
$\chi=2\pi\times4.6$ Hz \cite{Artur}.} \label{fig3}
\end{center}
\end{figure}

In Fig.~\ref{fig2}, we check the validity of Eq. (\ref{cma}) by
comparing it with the exact result of $|c_{m}|^{2}$ [where $c_{m}$
is given by Eq. (\ref{cm})], and find both results fit with each
other. For symmetric BEC with $\theta_0=\pi/2$, we have $\langle
\hat{J}_z\rangle=0$ and atom number variance $(\Delta
\hat{J}_{z})^{2}=j/2$; while for $\theta_0\neq\pi/2$, the variance
(width) becomes narrow, as shown in Fig.~\ref{fig2}(b). Inserting
Eq. (\ref{cma}) into Eq. (\ref{J+}), and replacing the discrete sum
over $m$ by an integral, we get
\begin{eqnarray}
\langle \hat{J}_{+}\rangle &\simeq &\frac{1}{\sqrt{\pi }}e^{i(\phi
_{0}+\delta t+\chi t)}e^{2i\left\langle
\hat{J}_{z}\right\rangle \chi t}\cot \frac{\theta _{0}}{2}  \nonumber \\
&&\times \int_{\rho _{\min }}^{\rho _{\max }}[j-\langle
\hat{J}_{z}\rangle -\sqrt{2}(\Delta\hat{J}_{z})\rho] \nonumber \\
&&\times \exp[-\rho ^{2}+2i\sqrt{2}\chi t(\Delta \hat{J}_{z})\rho]
d\rho ,  \label{J+a0}
\end{eqnarray}%
where we have set
\begin{equation}
\rho=\frac{m-\langle\hat{J}_{z}\rangle}{\sqrt{2}(\Delta
\hat{J}_{z})}.
\end{equation}%
For $\theta _{0}\sim \pi /2$ and $j\rightarrow \infty $ (large $N$
limit), the integral upper limit $\rho _{\max }=\sqrt{j}\tan (\theta
_{0}/2)\rightarrow \infty $ and the lower limit $\rho _{\min
}=-\sqrt{j}\cot (\theta _{0}/2)\rightarrow -\infty $. In the
short-time limit ($\chi t\ll 1$), $1-2i\chi t\cos ^{2}(\theta
_{0}/2)\simeq \exp \{ -2i\chi t\cos ^{2}(\theta _{0}/2)\}$, and thus
from Eq. (\ref{J+a0}) we obtain
\begin{eqnarray}
\langle \hat{J}_{+}\rangle &\simeq &j\sin (\theta_0)\exp \left\{
i\left[\phi_{0}+\delta t+\chi t(2j-1)\cos \theta _{0}\right] \right\}  \nonumber \\
&&\times \exp \left[-2\chi^{2}(\Delta\hat{J}_{z})^{2}t^{2}\right] ,
\label{J+a}
\end{eqnarray}%
where two integrals: $\int_{-\infty }^{\infty }e^{-\rho
^{2}}e^{i\lambda \rho }d\rho =\sqrt{\pi }e^{-\lambda ^{2}/4}$ and $
\int_{-\infty }^{\infty }\rho e^{-\rho ^{2}}e^{i\lambda \rho }d\rho
=i\lambda \sqrt{\pi }e^{-\lambda ^{2}/4}/2$ have been used in derive
of the above result. Thus, the short-time solution of the coherence
reads
\begin{equation}
g_{12}^{(1)}(t)\simeq \exp \left[ -2\chi ^{2}(\Delta
\hat{J}_{z})^{2}t^{2}\right] \equiv \exp \left[ -\left(
t/t_{d}\right) ^{2}\right] , \label{Imamoglu}
\end{equation}%
with characteristic time scale of the phase diffusion
\begin{equation}
\chi t_{d}=\frac{1}{\sqrt{2}(\Delta
\hat{J}_{z})}=\frac{1}{\sqrt{j}\sin \theta _{0}}.  \label{diffusion
time}
\end{equation}%
Obviously, the phase-diffusion time $\chi t_{d}=1/\sqrt{j}$ for the
initial CSS with $\theta_{0}=\pi/2$ \cite{Imamoglu,Vardi,Artur};
while for $\theta_{0}\neq \pi/2$, our results show that the
diffusion becomes slowly due to an enhanced factor $\sin\theta_{0}$
in the phase-diffusion rate [see also Fig.~\ref{fig3}]. The revival
of the coherence occurring later at a time $\chi t_{r}=\pi$
\cite{Castin,Vardi} is observable in real experiment, such as
$t_{r}=\pi/\chi\simeq 108.7$ microsecond for nonlinearity $\chi
\simeq 2\pi\times 4.6$ Hz and particle number $N=60$ \cite{Artur}.

%%%%%%%%%%%%%%%%%%%%%%%%%%%%%%%%%%%%%%%%%%%%%%%%%%
%%%%%%%%%%%%%%%%%%%%%%%%%%%%%%%%%%%%%%%%%%%%%%%%%%

Finally, we note that similar results of Eq.~(\ref{J+}) and
Eq.~(\ref{J+a}) have been obtained in Ref. \cite{Sinatra}. However,
the authors focused on the increase of the diffusion time as
$\chi\rightarrow 0$, but not $\theta _{0}\neq \pi/2$. In addition,
we emphasis that for negative $\chi$ the so-called phase separation
may take place due to dynamically unstable of the BECs
\cite{Chui,Kasa,MI1,MI2,MI3,MI4,MI5,MI6}. Both the phase diffusion
and the phase separation reduce the Ramsey fringes' visibility
(i.e., the first-order coherence $g_{12}^{(1)}$). However, the
latter effect results from dynamics of spatial degree-of-freedom,
rather than that of the internal states considered here.

In summary, we have investigated phase diffusion of two-component
BEC for arbitrary initial CSS. We show analytically that for the CSS
with $\theta _{0}\neq\pi /2$, the diffusion process is suppressed
due to the reduced atom number variance $(\Delta \hat{J}_{z})^{2}$
below the standard quantum limit $j/2$ [see also Eqs.
(\ref{variance}) and (\ref{diffusion time})]. Our analytic results
are based upon the SMA, which works well to describe the condensates
tightly confined in a three-dimensional (3D) trapping potential. For
a lower dimensional cases \cite{Yi}, the coherence $g_{12}^{(1)}$
[or the fringe visibility] is expected to decay more quickly due to
the whole continuum of excitation modes in confined degrees of
freedom \cite{Artur,Rafi}.

\section*{Acknowledgments}

This work is supported by the NSFC (Contract No.~10804007).

\end{document}